\documentclass[dvips]{article}
\usepackage{icrctc07}

\newcommand{\be}{\begin{equation}}
\newcommand{\ee}{\end{equation}}
\newcommand{\nn}{\mbox{} \nonumber \\ \mbox{} }
\newcommand{\ba}{\begin{eqnarray}}
\newcommand{\ea}{\end{eqnarray}}
\newcommand{\om}{\omega}
\newcommand{\Alfven}{ Alfv\'{e}n }

\newcommand{\E}{{\bf E}}
\newcommand{\B}{{\bf B}}

\newcommand\eg{\textit{e.g.\ }}

\newcommand{\Bf}{{magnetic field\,}}
\newcommand{\Bfs}{{magnetic fields\,}}
\newcommand{\Ef}{{electric  field\,}}
\newcommand{\Efs}{{electric fields\,}}
\newcommand{\CR}{{cosmic ray\,}}
\newcommand{\CRs}{{cosmic rays\,}}

\newcommand{\apj}{ApJ\,\,}

\newcommand{\physrep}{Phys. Rep.\,\,}

\title{Inductive acceleration of UHECRs in sheared relativistic jets}
\shorttitle{Inductive acceleration of UHECRs}
\authors{Maxim Lyutikov $^{1}$, Rachid Ouyed$^{2}$}
\shortauthors{Lyutikov and Ouyed}
\afiliations{$^{1}$Department of Physics, Purdue University, 525 Northwestern Avenue
West Lafayette, IN, USA\\ $^{2}$ Department of Physics and Astronomy, University of Calgary,
       2500 University Drive NW, Calgary, Alberta, T2N 1N4 Canada}
\email{lyutikov@purdue.edu}

\abstract{Relativistic outflows carrying large scale \Bfs have large 
inductive potential and may  accelerate protons to ultra high energies.
We discuss a novel scheme of Ultra-High Energy Cosmic
Ray (UHECR)  acceleration  due to drifts in 
 magnetized, cylindrically collimated, sheared jets of powerful active galaxies
(with jet luminosity $\geq 10^{46}$ erg s$^{-1}$).  A
 positively charged particle carried by such a plasma is in an unstable  
equilibrium if  ${\bf B} \cdot \nabla \times {\bf v}< 0$, 
so that kinetic drift along the velocity shear would lead
to fast, {\it regular} energy gain.  
The 
highest  rigidity  (ratio of energy to charge) particles  are 
accelerated most efficiently implying
 the dominance of light nuclei for energies above the ankle in our model: from a mixed population
of pre-accelerated particle the drift mechanism  picks up and boosts protons
preferably.}

\begin{document}
\maketitle
%Begin the section.

\section{Introduction}

There is a consensus that relativistic outflows, and  AGN jets in
particular, are accelerated to relativistic speeds and collimated by
large scale \Bfs threading accretion disk and central black hole.
 Energetically, \Bfs may carry a large 
fraction of jet luminosity \cite{bla02,lb03}. At largest scales \Bf is dominated
by a toroidal component $B_\phi$. In addition,
  toroidal \Bf may provide jet collimation, so that asymptotically AGN
jets may also be fully collimated to a cylindrical shape \cite{HeyN03}.  
Axial motion of toroidal \Bf create radial (in cylindrical coordinates) \Ef.
One also expects that  jets are sheared, so that the 
central spine of the jet is moving with larger velocity than
its periphery. In a sheared jet Lorentz transformation cannot get rid of \Ef everywhere in space.

Powerful astrophysical outflows carrying large scale \Bfs poses large inductive potentials that may be able to accelerate UHECRs \cite{lo07}. If the Poynting luminosity of a source is $L_{EM}$, then the total inductive potential is 
\be
  \Phi \sim \sqrt{ 4 \pi  L_{EM} \over c} =
  4 \times 10^{20} {\rm V} \left({ L \over 10^{46} {\rm erg/sec}}
  \right)^{1/2}
   \label{phi}
  \ee 
  Though Poynting luminosity  $L_{EM}$ may exceed the observed total luminosity of a source (which is related to the rate of dissipation), the estimate (\ref{phi})
 {\it excludes acceleration of UHECRs in low power
AGNs  (\eg Cen A and  M87), 
low power BL Lacs and starburst galaxies (\eg M82 \& NGC 253)}  and  limits the possibilities to more distant high power AGNs like 
 higher power FR I, 
FR II radiogalaxies, radio-loud quasars  and GRBs.

Another constraint that acceleration sites should satisfy is that
radiative losses should not degrade particle energy. We can derive very general constraints
on possible location of \CR acceleration just by balancing the
most efficient acceleration, by $E\sim B$, and 
radiative losses. f we normalize total energy density to energy density of \Bf
$u= \zeta u_B$, $\zeta > 1$, we find
\be
{\cal E}  = mc^2 \sqrt{ c  \over \zeta r_c Z^3 \om_B}  
\label{2}
\ee
where $r_c = e^2/m c^2$. 
This gives limits  on \Bf and the distance from the central source
 \ba &&
B < { m^2 c^4 \over \zeta  Z^3 e^3}  \left( { m c^2 \over {\cal E} } \right)^2 \Gamma^3
= 
\nn &&
2 \, \Gamma^3 \, {\rm G} \,  \left( { {\cal E} \over 100 EeV} \right)^{-2}  
\left( {1 \over Z} \right)^3
\nn &&
R >   {Z^2 e^2 \zeta  \over m c^2}  \left( { {\cal E} \over m c^2} \right)^3 { 1 \over \Gamma^2} =
\nn &&
 10^{17} \, { 1 \over \Gamma^2} \, {\rm cm} \,
 \left( { {\cal E} \over 100 EeV} \right)^3  \left( Z \right)^2
 \label{B}
\ea
 Where we also allowed a possibility that plasma is expanding with bulk Lorentz factor
 $ \Gamma$, so that \Bf in the plasma rest-frame  is $B/\Gamma$ and
 typical scale is $R/\Gamma$, and assumed $\zeta=1$; $EeV= 10^{18}$ eV. 

Relations (\ref{B}) show that higher energy \CRs are better accelerated at larger distances.
 AGN jets, which propagate
to more than  100 kpc distances present an interesting possibility. Note, that as long as the jet remains relativistic, the total inductive potential is approximately  conserved, 
so one can  ``wait'' a long time for a particle to get accelerated without worrying about radiative loses.  Thus, UHECRs can be accelerated  inside the jet at distances from a fraction
of  a parsec (Eq. (\ref{B})) to hundreds of kpc, as long as the jet remains relativistic
and sustains a large inductive potential.

Since inductive \Efs are orthogonal to \Bf, particles cannot move freely along them. Thus, 
it is not obvious how to achieve energy gain.
One possibility is kinetic drift  which may result in   regular motion  across \Bf and along \Ef
leading to  {\it regular energy gain} as compared
to the stochastic, Fermi-type schemes.
 Since the  drift velocity  increases with particle energy, {\it the rate of energy gain 
will also  increases with particle energy}. Thus, the highest energy
particles  will be accelerated most efficiently. This means that 
from a pre-accelerated population, the mechanism proposed
here will pick up particles with highest
energy and will boost them to even  higher values.
In addition, when a particle has crossed a considerable fraction
of the total available potential,
it gyro-radius becomes comparable with flow scale.
In this case, drift approximation brakes down. As a result,
as  long as  the  particle remains inside a jet,
{\it the  acceleration rate reaches the
theoretical maximum of an inverse gyro-frequency}.
Thus, the most efficient acceleration occurs right before the particle leaves the jet.
One may say that
 in case of inductive acceleration, {\it becoming unbound is beneficial
 to acceleration}, contrary to the case of stochastic acceleration when
 for unbound particles acceleration ceases.

\section{Particle dynamics in sheared flow}
\label{motion}

\begin{figure}[ht]
\includegraphics[width=0.9\linewidth]{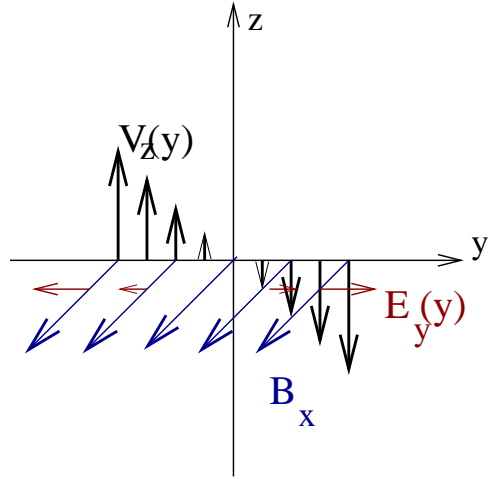}
\caption{Electric field in a sheared flow in a frame where surface the $y=0$
is at rest.  For $\B \cdot \nabla \times {\bf v} < 0$
 the \Ef is directed away from the $y=0$ surface. }
\label{shearCR}
\end{figure}

In a
transversely sheared flow one sign of charges is located at a maximum of electric
potential, as we describe in this section. Consider 
 sheared flow carrying \Bf. At each point there is electric field $\E = - {\bf v} \times \B/c$, so that the 
electric potential is determined by
\be
\Delta \Phi = { 1 \over c}  \nabla \cdot \left({\bf v} \times \B  \right)
\label{Phii}
\ee
In a local rest frame, where ${\bf v} =0$,  this becomes
vanishes at the position of a particle,
\be
\Delta \Phi = { 1 \over c} \left( \B \cdot \nabla \times {\bf v} \right)
\label{Phi}
\ee
Thus, {\it  depending on the sign of the quantity 
$ \left( \B \cdot \nabla \times {\bf v} \right)$ (which is a scalar) charges  of 
one  sign are  near potential minimum, while those with the
opposite sign are near  potential maximum}, see Fig. \ref{shearCR}.
Since electric field is perpendicular both to velocity and magnetic field, locally, the
electric potential is a function of only one coordinate along this direction.
For $ \left( \B \cdot \nabla \times {\bf v} \right) <0 $ (we will call this case negative shear)
ions are near potential maximum. 

 Inductive \Efs are not easily accessible for acceleration
since particle need to move across \Bf, a processes prohibited under 
ideal Magneto-Hydrodynamics (MHD) approximation.  On the other hand, 
kinetic drifts may result in   regular motion  across \Bf and along \Ef
leading to  {\it regular energy gain}. Typically, the direction of drift is along the normal to the \Bf and to the
direction of the force that induces a drift.
 In sheared cylindrical jet with toroidal \Bf, the 
electric field is in radial direction, so that 
 in order to gain energy particle should  experience 
radial drift. It is then required that there should be a force along the
axis. Such force may arise if   a jet is axially  inhomogeneous, \eg due to propagation of compressible waves along the jet,  
resulting in gradient drift due to changing magnetic pressure.

Let's assume that there is a long wavelength inertial \Alfven wave propagating
along the $z$-direction with a phase speed $V_A$. For $V_A \ll c$, 
 the magnetic perturbation  $\delta B$ 
in the wave is much larger than electric perturbations $\delta E$
by a factor $c/V_A$, so that the wave is nearly magneto-static
(in other words $ \om \ll k_{z} c$). A test particle will experience a drift
in the $x$-direction with magnitude
\be
u_d \sim {\delta B \over B_0} { \gamma c^2 k_{z} \over Z \om_{B,0} } \sim { \gamma c^2 k_{z} 
\over Z \om_{B,0}} 
\ee
where we assumed strong perturbation $\delta B_0 \sim  B$ 
and $\om_{B,0} = e  B_0/mc$.
As a particle drifts along \Ef in the $x$-direction its Lorentz factor 
evolves according to
\be
\partial _ t  \gamma = {Z e E_{y} u_d  \over m c^2} =   \beta_0{ Z \om_{B,0}  u_d^2 t \over c L_V}
\sim \beta_0  { \gamma^2 c^3 k_{z}^2 t \over Z \om_{B,0}  L_V}
\label{q1}
\ee
Thus, energy gain or loss of a test particle depends
on direction of \Bf,  sign of charge (through $\om_B$) and 
direction of velocity vorticity (through sign of $L_V$):
in other words it depends on sign of shear. On the
other hand, it is {\it independent} of the direction
of the drift.

Since drift velocity increases with particle energy, the rate of energy gain
also increases with energy, see Eq. (\ref{q1}). This leads to 
one of the most unusual properties of the proposed acceleration mechanisms:
 {\it  highest energy (or highest rigidity)
 particles are accelerated most efficiently}. In addition,
 at the last stages of acceleration, when particle
Larmor radius becomes of the order of jet scale, particle motion in positive shear flow
becomes unstable even without gradient drift while  {\it  acceleration rate
does reach absolute theoretical maximum of inverses relativistic gyro-frequency}.

In addition, since acceleration rate is {\it inversely}
proportional to charge,  
at a given energy small charge (higher rigidity) particles
are accelerated most efficiently. Thus, from a population 
of pre-accelerated particles with mixed composition, drift mechanism
will pick up particles with smallest charge: protons.
This explains why above the ankle protons start to dominate over heavy
nuclei.

We can also calculate evolution of the spectrum. 
If the  initial injection spectrum is power-law, $f_0 \propto 1/\gamma_0^p$, then
\begin{eqnarray}
f(\gamma,t) &\propto&
{ 1\over \gamma ^p} \left(1+\gamma \left(  {t\over \tau_0}\right)
^2 \right)^{p-2} \\\nonumber
&\sim&  \gamma ^{-2} , \quad \mbox{for $\gamma t^2/\tau_0^2 \gg 1$}
\label{f}
\end{eqnarray}
Thus, for $p>2$  the spectrum flattens  with time.

The hardest spectrum that can be achieved has a power law index of 2.
This limiting case corresponds to unlimited acceleration in a  plane-parallel geometry,
which is realistically applicable to energies well
below the total available potential. At highest energies the final
spectrum will depend on the distribution of pre-accelerated particles with
respect to the electric potential and, in case of contribution from many
sources, on distribution of total potentials.

\section{Discussion/predictions}
\label{discuss}

The best astrophysical  location for operation of the proposed mechanism is 
{\it cylindrically collimated}, high power AGN jets. The proposed mechanism cannot work
in spherically (or conically) expanding outflows since in this case 
 a particle  experiences polarization drift, which is a {\it first order} in Larmor radius,
 due to the fact  that in the flow frame magnetic field decreases with time. 
For a constant velocity flow this drift is always against the electric field (for a positively
charged particle) and lead to the decrease of energy on time
scale  $R/(c \Gamma)$, where $R$ is  a distance from the central source
and $\Gamma$ is the Lorentz factor of the flow. 
Thus, in spherically expanding flows adiabatic losses always dominate over
regular energy gain due to drift motion: the proposed mechanism 
would fail then.  On theoretical grounds,
AGN jets (or at least their cores) may indeed be asymptotically 
cylindrically collimated  \cite{HeyN03}. 
Observations of large scale jets, \eg Pictor A, do show jets that seem to be cylindrically
collimated on scales of tens of kiloparsec.

In cylindrically collimated parts of the jet acceleration can happen from sub-parsec to hundreds of
kiloparsec scales: as long as the motion of the jet is relativistic the total electric potential
remains approximately constant. Thus, UHECRs need not be accelerated close
to the central black hole where radiative losses are important.
After a jet has propagated  parsecs from the central source
 radiative losses become negligible.

Another  constraint on the mechanism comes from the requirement that in order
 to produce radial kinetic drift the 
jet \Bf should be inhomogeneous along the axis. Though shocks provide
 possible inhomogeneity of \Bf ($\delta$-function inhomogeneity on the shock
front), we disfavor shock since particles are advected downstream and cannot
drift large distances along shock surface.
% (for relativistic shocks, which are generically 
%quasi-transverse, it is then required that scattering rate be super-Bohm).
In a gradual inhomogeneity a particle drifts orthogonally to the 
field gradients and thus generally will remain in the region of 
inhomogeneous fields.
Extragalactic jets are expected to have axial inhomogeneities,  both due to non-stationary 
conditions at the source and 
due to propagation of compression and rarefaction waves generated
at the jet boundary via interaction with surrounding plasma.

Our model has a number of 
clear predictions, some of which are related to astrophysical  association of acceleration sites of UHECRs with AGN jets \cite{olinto00} and some are specific to the model:
  (i)  one needs a relatively powerful AGN,
with luminosity $\geq 10^{46}$ erg/sec. This limits possible sources
to high power sources like FR II radiogalaxies,  radio loud quasars and high power BL Lacs
(flat spectrum radio quasars).
  Powerful AGNs are relatively rare and far
apart, so that a  steep GZK cut-off  corresponding to large source separation should be seen. 
(ii)   UHECRs come from sources with low spacial density. This may be reflected in the
distribution of arrival directions. 
(iii) Extragalactic  UHECRs   should  be  dominated by  protons.
(iv) Depending on ''extra-galactic seeing conditions'' arrival directions of  UHECRs may point to
their sources, though complicated \Bf structure may erase this correlation.
In addition,
the fact that only flows with negative shear can accelerate protons
implies that only approximately half of such AGNs can be sources of UHECRs (this assumes that the AGN central engine - black hole or an accretion disk - is  dominated by  large scale, dipolar-like \Bf).

%\begin{references}

%\bibliography{iCRC07-1}
%This in the bibtex style, is ok.
%\bibliographystyle{plain}

\end{document}